\documentstyle[12pt]{article}

\newcommand{\R} [1] {~\mbox{(\ref{#1})}}

\begin{document}

\title{Capillary-gravity waves: The effect of viscosity \\
on the wave resistance}
\author{D. Richard, E. Rapha\"el\\
Coll\`ege de France\\
Physique de la Mati\`ere Condens\'ee \\
URA 792 du CNRS\\
11, place Marcelin Berthelot,\\
 75231 Paris cedex 05, France.}

\maketitle

\begin{abstract}
The effect of viscosity on the wave resistance experienced by a
two-dimensional perturbation moving at uniform velocity over the
free surface of a fluid  is investigated. The analysis is based on
Rayleigh's linearized theory of capillary-gravity waves. It is shown in
particular that
the wave resistance remains {\it bounded} as the velocity of the
perturbation approaches the minimum phase speed $c^{min} = (4 g
\gamma  /\rho)^{1/4}$ ($\rho$ is the liquid density, $\gamma$ is
the liquid-air surface tension, and $g$ the acceleration due to
gravity), unlike what is predicted by the inviscid theory.
\end{abstract}
\vspace{1.5cm}

\newpage
\par Consider a body of liquid in equilibrium in a gravitational
field and having a planar free surface. If, under the action of some
external perturbation, the surface is moved from its equilibrium
position at some point, motion will occur in the liquid. This
motion will be propagated over the whole surface in the form of
waves, which are called {\it capillary-gravity waves}
\cite{LandauLifshitz}. These waves are driven by a balance between the
liquid inertia and its tendency, under the action of gravity
and under surface tension forces, to return to a state of stable
equilibrium. For an inviscid liquid of infinite depth,
the relation between the circular frequency $\omega$
and the wave number $k$ (i.e., the {\it  dispersion relation})
is given by ${\omega}^2 = g k + \gamma k^3/\rho$
where $\rho$ is the liquid density, $\gamma$
the liquid-air surface tension, and $g$ the acceleration due to
gravity \cite{Acheson}. The above equation may also
be written as a dependence
of wave speed $c = \omega/k$ on wave number
%%%%%%%%%%%%%%%%%%%%%%%%%%%%%%%%%%%%%%%%
\begin{equation}
\label{relationdispersion}
c = {\left(g/k + \gamma k/\rho\right)}^{1/2}
\end{equation}
%%%%%%%%%%%%%%%%%%%%%%%%%%%%%%%%%%%%%%%%
An important feature of Eq.\R{relationdispersion} is
that it implies a minimum phase speed of
$c^{min} = (4 g \gamma  /\rho)^{1/4}$ reached at $k_{min} = \kappa$
where $\kappa^{-1} = {\left[\gamma/(\rho g)\right]}^{1/2}$
is the capillary length \cite{Rayleigh}.
For water with $\gamma = 73 \; \rm{mN \; m^{-1}}$
and $\rho = 10^3 \; \rm{kg \; m^{-3}}$, the minimun phase
speed is $c^{min} = 0.23 \; \rm{m \; s^{-1}}$
and the corresponding wavelength is
$\lambda^{min} = 2\pi/\kappa = 1.7 \;10^{-2} \; \rm{m}$.
The dispersive property of capillary-gravity waves is
responsible for the complicated wave pattern generated at the free
surface of a still liquid by a disturbance moving with a velocity
$V$ greater than  $c^{min}$ \cite{Acheson}. The disturbance
may be produced by a small
object (such as a fishing line) immersed in the liquid or by the
application of an external surface pressure distribution
$P_{ext}$. The waves generated by the moving disturbance
continuously remove energy to infinity. Consequently, for $V >
c^{min}$, the disturbance will experience a drag, $R$, called the
{\it wave resistance} \cite{Lighthill}. For $V < c^{min}$, the wave
resistance is equal to zero since, in this case, no waves are
generated by the disturbance. A few years ago \cite{RaphaeldeGennes},
it has been predicted that the wave resistance corresponding to a
surface pressure distribution symmetrical about a point should be
{\it discontinuous} at $V = c^{min}$ \cite{Webster}. This prediction
has been checked very recently by Browaeys and
co-workers using a magnetic fluid \cite{Browaeys}. The experimental
results of Browaeys {\it et al.} indicate, however, that the
disturbance experienced a small but nonzero drag for $V < c^{min}$.
Since this nonzero drag might be due, in part, to the finite
viscosity of the fluid, it is of some importance to incorporate
this physical parameter in the inviscid model of
Ref.\cite{RaphaeldeGennes}. In order to simplify the discussion, we
will here consider the case of a pressure distribution $P_{ext}$
localized along a line. The more complicated case of an
axisymmetric surface pressure distribution will be consider
elsewhere.
\par We take the $xy$-plane as the equilibrium surface of the
fluid and assume that a pressure distribution of the form
%%%%%%%%%%%%%%%%%%%%%%%%%%%%%%%%%%%%%%%%
\begin{equation}
\label{pressurfield}
P_{ext} = P_0 \frac{b}{\pi (b^2+x^2)}
\end{equation}
%%%%%%%%%%%%%%%%%%%%%%%%%%%%%%%%%%%%%%%%
travels over the surface in the
$x$-direction with a velocity $V$ (in all
what follows we assume that $b \ll \kappa^{-1}$).
It can be shown that in the  case on an inviscid liquid
the wave resistance per unit length corresponding to the external
pressure distribution\R{pressurfield}
is given by \cite{Lighthill,RaphaeldeGennes}:
%%%%%%%%%%%%%%%%%%%%%%%%%%%%%%%%%%%%%%%%
\begin{equation}
\label{resistanceinviscid}
R = \frac{P_0^2}{\gamma (k_1-k_2)} \left[k_1 e^{-2 b k_1} + k_2 e^{-2 b
k_2}\right] \qquad (V > c^{min})
\end{equation}
%%%%%%%%%%%%%%%%%%%%%%%%%%%%%%%%%%%%%%%%
(remember that for an inviscid liquid, $R = 0$
for $V < c^{min}$). In Eq.\R{resistanceinviscid},
the wave numbers $k_1$ and $k_2$ denotes the two (real) solutions of $(g/k +
\gamma k/\rho)^{1/2} = V$ (see Eq.\R{relationdispersion}).
A brief inspection of Eq.\R{resistanceinviscid} shows
that the wave resistance is a decreasing function of the
perturbation velocity $V$.
In the limit $V \gg c^{min}$, Eq.\R{resistanceinviscid} reduces to
$R \simeq (P_0^2/\gamma) e^{-4 (b/\kappa^{-1}) (V/c^{min})^2}$ .
As the velocity $V$ decreases towards $c^{min}$, the wave resistance
Eq.\R{resistanceinviscid} becomes unbounded.
This result is directly related to the fact that as
$V$ approaches $c^{min}$, the energy transferred by the moving
pressure distribution cannot be radiated away.

We now turn our attention to the case of a viscous liquid and
investigate how the wave resistance\R{resistanceinviscid} is
modified by the liquid viscosity. In order to calculate $R$, we may
imagine a rigid cover fitting the surface everywhere, as suggested
by Havelock \cite{Havelock}. The assigned pressure system  $P_{ext}$
is applied to the liquid surface by means of this cover; hence, the
wave resistance is simply the total resolved pressure per unit
length in the $x$-direction \cite{Havelock}:
%%%%%%%%%%%%%%%%%%%%%%%%%%%%%%%%%%%%%%%%
\begin{equation}
\label{Havelock}
R = - \int P_{ext}(x) \left(\frac{d}{dx} \zeta(x) \right) \, dx
\end{equation}
%%%%%%%%%%%%%%%%%%%%%%%%%%%%%%%%%%%%%%%%
where $\zeta (x)$ denotes (in the frame of the perturbation)
the displacement of the free surface
from its equilibrium position.
Let $\hat{\zeta }$ and ${\hat{P}}_{ext}$
denote the Fourier transforms of $\zeta$ and ${P}_{ext}$,
respectively. Using the Navier-Stokes equation
for a viscous fluid along with the
appropriate stress condition at the free surface \cite{Lamb},
one can relate $\hat{\zeta }$ to ${\hat{P}}_{ext}$
through the relation

%%%%%%%%%%%%%%%%%%%%%%%%%%%%%%%%%%%%%%%%
\begin{equation}
\label{HaveHave}
\left[ (2 \nu k^2 - i V k)^2 + g \mid \hspace*{-1.2mm} k \hspace*{-1.2mm} \mid +
\frac{\gamma}{\rho} {\mid \hspace*{-1.2mm} k \hspace*{-1.2mm} \mid}^3 -
4 \nu^2 {\mid \hspace*{-1.2mm} k \hspace*{-1.2mm} \mid}^3 \sqrt{k^2 -
i \frac{V}{\nu} k} \;\right] \hat{\zeta } =
- \frac{\mid \hspace*{-1.2mm} k \hspace*{-1.2mm} \mid}{\rho} {\hat{P}}_{ext}
\end{equation}
%%%%%%%%%%%%%%%%%%%%%%%%%%%%%%%%%%%%%%%%
where the parameter $\nu$ is the kinematic viscosity of the liquid.
Inserting Eq.\R{HaveHave} into Eq.\R{Havelock} we obtain
%%%%%%%%%%%%%%%%%%%%%%%%%%%%%%%%%%%%%%%%
\begin{equation}
\label{resistance}
R = \frac{P_0^2}{ \pi \gamma} \quad \Re \left( \int\limits_{0}^{\infty}
\frac{i k^2 \;\exp{\left(-4 \frac{b }{\kappa^{-1}}v^2\;k\right)}}{(2 \epsilon_0 \;v\; k^2-i k)^2
+\frac{1}{4} v^{-4} k
+k^3-4 \epsilon_0^2\; v^2 \;k^3 \sqrt{k^2-i \frac{k}{\epsilon_0\;v}}} dk
\right)
\end{equation}
%%%%%%%%%%%%%%%%%%%%%%%%%%%%%%%%%%%%%%%%
where $v = V/c^{min}$ and $\epsilon_0 = \eta c^{min}/\gamma$ . The
symbol $\Re$ represent the real part of a complex expression. The
wave resistance Eq.\R{resistance} is shown graphically in figure (1) as a function of
the reduced velocity $v$ for $\epsilon_0 = 3 \; 10^{-3}$
and $b=2.5 \; 10^{-3} \;
\kappa^{-1}$
(the inset  highlights the behaviour of $R$
at low velocities).
Two important features should be noted in comparison
to the inviscid case:
(a) First, while $R$ increases steeply near $V = c^{min}$,
it remains bounded; (b) Secondly, as soon as the perturbation
velocity $V$ is greater than zero, the wave resistance takes
finite values. This late result is a direct consequence
of the internal viscous dissipation inside the liquid.
%%%%%%%%%%%%%%%%%%%%%%%%%%
\par For $\epsilon_0$ much smaller than unity,
Eq.\R{resistance} can be simplified and
the above two features can be recovered analytically
(for water with $\gamma = 73 \; \rm{mM \; m^{-1}}$
and $\rho = 10^3 \; \rm{kg \; m^{-3}}$,
$\epsilon_0 \approx 3 \; 10^{-3} \ll 1$).
Using standard mathematical technics \cite{Morse},
one can show that the wave resistance displays a maximum of

\begin{equation}
\label{max}
R =R_{max} \; \approx \frac{P_0^2}{\gamma \; \sqrt{\epsilon_0}}
\end{equation}

\noindent for
\begin{equation}
\label{V_max}
V = V_{max} \; \approx c^{min}\,(1+\epsilon_0)
\end{equation}
\\
On the other hand, in the limit $V \ll c^{min}$,
the wave resistance $R$ varies linearly
with the perturbation velocity:
%%%%%%%%%%%%%%%%%%%%%%%%%%%%%%%%%%%%%%%%
\begin{equation}
\label{Stokes}
R \approx \frac{4 P_0^2}{\pi\gamma}\,\epsilon_0\left(\frac{V}{c^{min}}\right)
\,{\rm Log}\left(\frac{\kappa^{-1}}{b}\right)
\end{equation}
%%%%%%%%%%%%%%%%%%%%%%%%%%%%%%%%%%%%%%%%

\noindent This linear behavior can be observed in the insert in figure (1).
%%%
\par Let us conclude by a few remarks.
In the calculations made above for the wave resistance $R$, we
have used Rayleigh's linearized theory of capillary-gravity waves
\cite{Rayleigh}. We have shown that one of the effects of
viscosity was to cutoff the unbounded response of the liquid
predicted by the inviscid model as $V \searrow c^{min}$. For a
small viscosity, the response of the liquid remains, however,
rather large near $c^{min}$ (see figure (1)), and it would be of
some interest to take nonlinear effects into account in the
calculation of $R$. This is beyond the scope of the present
letter. For a recent review of nonlinear capillary-gravity waves,
the reader is referred to the work of Dias and Kharif \cite{Dias}.
In the present study we have emphasized the asymptotic
behavior of the wave resistance for a liquid of low viscosity.
We hope to explore the high
viscosity limit in a subsequent report.
Note also that the calculations presented in this letter assumed
an external pressure distribution localized along a band. Further
work will assess the effect of viscosity for a pressure field
symmetrical around a point. \vspace{1cm}
\\
{\bf Acknowledgements.} This work was motivated by discussions
with P.-G. de Gennes. We would like to thank him as well as J.-C. Bacri, J.
Browaeys, F. Dias and D. Quéré for helpful comments. \vspace{1cm}
\\
{\bf Figure caption : } $R^* = \pi\gamma \; P_0^{-2} R $ as a
function of $v=V/c^{min}$, with R being the wave resistance Eq.
\R{resistance}, $\epsilon_0 = 3 \; 10^{-3}$ and $b=2.5 \;
10^{-3} \; \kappa^{-1}$. The inset shows the behavior at low
velocities. \vspace{1cm}
\\

\end{document}